\def\ep{E_{\rm peak}}
\documentclass[onecolumn]{pasj01}
\draft 
\Received{$\langle$reception date$\rangle$}
\Accepted{$\langle$acception date$\rangle$}
\Published{$\langle$publication date$\rangle$}

\begin{document}

\title{Detection of the thermal component \\in GRB 160107A}
\author{Yuta \textsc{Kawakubo}\altaffilmark{1,}$^{*}$, Takanori \textsc{Sakamoto}\altaffilmark{1}, Satoshi \textsc{Nakahira}\altaffilmark{2}, \\Kazutaka \textsc{Yamaoka}\altaffilmark{3}, Motoko \textsc{Serino}\altaffilmark{1}, 
Yoichi \textsc{Asaoka}\altaffilmark{9},
M.~L. \textsc{Cherry}\altaffilmark{10}, 
Shohei \textsc{Matsukawa}\altaffilmark{1}, 
Masaki \textsc{Mori}\altaffilmark{4}, 
Yujin \textsc{Nakagawa}\altaffilmark{5}, \\
Shunsuke \textsc{Ozawa}\altaffilmark{9}, 
A.~V. Penacchioni\altaffilmark{7,8}, 
S.~B. \textsc{Ricciarini}\altaffilmark{6}, 
Akira \textsc{Tezuka}\altaffilmark{1}, \\
Shoji \textsc{Torii}\altaffilmark{9}, 
Yusuke \textsc{Yamada}\altaffilmark{1}, and 
Atsumasa  \textsc{Yoshida}\altaffilmark{1}}
\altaffiltext{$^1$}{Department of Physics and Mathematics, Aoyama Gakuin University, \\5-10-1 Fuchinobe, Chuo-ku, Sagamihara, Kanagawa 252-5258, Japan}
\altaffiltext{$^2$}{MAXI team, RIKEN, 2-1 Hirosawa, Wako, Saitama 351-0198, Japan}
\altaffiltext{$^3$}{ Institute for Space-Earth Environmental Research (ISEE), Nagoya University, \\ Furo-cho, Chikusa-ku, Nagoya, Aichi 464-8601, Japan}
\altaffiltext{$^4$}{Department of Physical Sciences, College of Science and Engineering, Ritsumeikan University, 1-1-1 Noji-higashi, Kusatsu, Shiga 525-8577, Japan}
\altaffiltext{$^5$}{Center for Earth Information Science and Technology, Japan Agency for Marine-Earth Science and Technology, 3173-25 Showa-machi, Kanazawa-ku, Yokohama, Kanagawa 236-0001, Japan}
\altaffiltext{$^6$}{Institute of Applied Physics (IFAC), National Research Council (NRC), Via Madonna del Piano, 10, 50019 Sesto, Fiorentino, Italy}
\altaffiltext{$^7$}{Department of Physical Science, Earth and Environment, University of Siena, via Roma 56, 53100, Siena, Italy}
\altaffiltext{$^8$}{ASI Science Data Center (ASDC), Via del Politecnico snc, 00133 Rome, Italy}
\altaffiltext{$^9$}{Research Institute for Science and Engineering, Waseda University, 3-4-1 Okubo, Shinjuku, Tokyo 169-8555, Japan}
\altaffiltext{$^{10}$}{Department of Physics and Astronomy, Louisiana State University, 202 Nicholson Hall, Baton Rouge, LA 70803, USA}

\email{ykawakubo@phys.aoyama.ac.jp}


\maketitle

\begin{abstract}
We present the detection of a blackbody component in GRB 160107A emission 
by using the combined spectral data of the CALET Gamma-ray Burst Monitor (CGBM) 
and the MAXI Gas Slit Camera (GSC).   The MAXI/GSC detected the emission 
$\sim$45 s prior to the main burst episode observed by the CGBM.  
The MAXI/GSC and the CGBM spectrum of this prior emission period is well fit 
by a blackbody with the temperature of $1.0^{+0.3}_{-0.2}$ keV plus a power-law with the photon 
index of $-1.6 \pm 0.3$.   We discuss the radius 
to the photospheric emission 
and the main burst emission based on the observational properties.  We stress 
the importance of the coordinated observations via various instruments collecting 
the high quality data over a broad energy coverage in order to understand the GRB prompt 
emission mechanism.  
\end{abstract}

\section{Introduction}
Gamma-ray Bursts (GRBs) are short (from a few ms up to $\sim$100 seconds) 
and intense flashes of gamma-rays.  GRBs are classified into short GRBs 
and long GRBs depending on whether their durations are shorter or longer than 2 s 
\citep{kouveliotou1993}.  It is widely accepted that these two GRB populations 
originate from different progenitors.  The long GRBs are the result of explosions of 
massive stars (e.g., \cite{woosley2006}), while the short GRBs are due to mergers 
of two neutron stars or a neutron star -- black hole binaries
\citep{eichler1989,paczynski1991,narayan1992}.  
Merger events have been shown to be gravitational wave sources detectable 
by the current generation detectors such as the Laser Interferometer 
Gravitational-Wave Observatory (LIGO) and the Virgo \citep{LIGO2016,MMA2017}.  

The radiation processes for the prompt GRB emission are still far from 
being understood.  Most of the GRB prompt emission spectra are well represented by 
a smoothly connected broken power-law function -- Band function \citep{Band_1993}.  
The break energies, $\ep$ (the peak energy in the $\nu F_{\nu}$ spectrum), are broadly 
distributed from a few keV up to a few MeV range \citep{sakamoto2005}.  This broad $\ep$ distribution 
reflects the intrinsic properties of GRB spectra (e.g., \cite{amati2002}); its origin 
is not clear.  Although the observed spectrum is due to the Synchrotron radiation 
from the relativistic electrons (e.g., \cite{tavani1996}), the low-energy photon indices of some of the observed spectra 
show a harder index than expected from the Synchrotron radiation (e.g., \cite{preece1998}).  
A unique observation is needed to break through the current situation.  

One interesting characteristic seen in some GRBs is the emergence of thermal emission 
in the prompt GRB spectra (e.g., \cite{Frontera_2001}).  The spectrum of the prompt emission of GRB~041006 shows blackbody 
components superimposed on the non-thermal spectrum \citep{shirasaki2008} with time-averaged temperatures 0.2 and 0.4 keV.  
\citet{starling2012} identified a blackbody component 
with the temperature between 0.1 and 1 keV in the {\it Swift} data for nine GRBs.  Identifications of the blackbody 
component with temperatures of several 10 keV up to 100 keV are claimed by the BATSE and the 
{\it Fermi}/GBM data \citep{ryde2005,ryde2010,guiriec2013}.  Although thermal emission from the photosphere 
is expected in the standard fireball model \citep{goodman1986,paczynski1986}, the wide dynamic range of the 
claimed blackbody temperature distribution in the spectra observed in prompt GRB is difficult to understand 
in a simple way.  

The CALorimetric Electron Telescope (CALET) was attached to the Exposed Facility of the Japanese Experimental Module (JEM) on 
the International Space Station (ISS) in August 2015 \citep{Torii_2011}.  The main scientific goals of 
CALET are the observations of high energy cosmic electrons and high energy GeV-TeV gamma-rays 
thanks to its thick calorimeter.  To support the gamma-ray observations, 
CALET includes CALET Gamma-ray Burst Monitor  (CGBM) to observe the prompt emissions of GRBs \citep{Yamaoka_2013}.  
After a month of initial checkout, CGBM began its scientific observation on October 2015.  
CGBM consists of two kinds of scintillation detectors to cover a wide energy range from 7 keV up to 
20 MeV:  1) The Hard X-ray Monitor (HXM) consists of a LaBr$_{3}$(Ce) crystal (61 mm diameter 
and 1.27 mm thickness) with 410 $\mu$m thickness Be window and a photomultiplier tube (PMT) 
covering the energy range from 7 keV to 1 MeV.  There are two identical HXM units (HXM1 and HXM2) 
facing at the same direction roughly 10$^{\circ}$ tilted from the zenith direction.  2) The other 
scintillator is the Soft Gamma-ray Monitor (SGM), 
utilizing a BGO crystal (102 mm in diameter and 76 mm in thickness) and a PMT to cover the energy 
range from 40 keV to 20 MeV.  The SGM points to the zenith direction.  The field of 
views of HXM and SGM are $\sim$120$^{\circ}$ 
and $\sim$2$\pi$ sr.   The GRB detection rate of CGBM is $\sim$4 GRBs per month.  

The Monitor of All-sky X-ray Image (MAXI) is also mounted on the Exposed Facility of the JEM-ISS \citep{Matsuoka_2009}.  
MAXI consists of two different slit cameras: Gas Slit Camera (GSC) and Solid-state Slit Camera (SSC) consisting 
of 1 dimensional position-sensitive proportional counters and X-ray CCDs respectively.    
GSC covers the energy range of 2 -- 30 keV and the two instantaneous fields of view of 
$1^{\circ}.5$ $\times$ 160$^{\circ}$ at the zenith and the horizontal direction of the ISS.  Because 
CALET and MAXI sit on the same platform, most of the MAXI/GSC and CGBM fields of view overlap.  

In this paper, we report on the prompt emission properties of GRB 160107A.  In subsection 2.1, 
the temporal results from the CGBM and the MAXI/GSC data, and also the localization 
by the MAXI/GSC and the Interplanetary network, are shown.  We present the CGBM and 
the MAXI/GSC spectral properties in subsection 2.2.  We discuss our result in section 3.  
The quoted errors are at the 90\% confidence level unless stated otherwise.  

\section{Observation and Data analysis}

\subsection{Temporal Properties and Localization}

\begin{figure*}[h]
 \begin{center}
  \includegraphics[width=100mm]{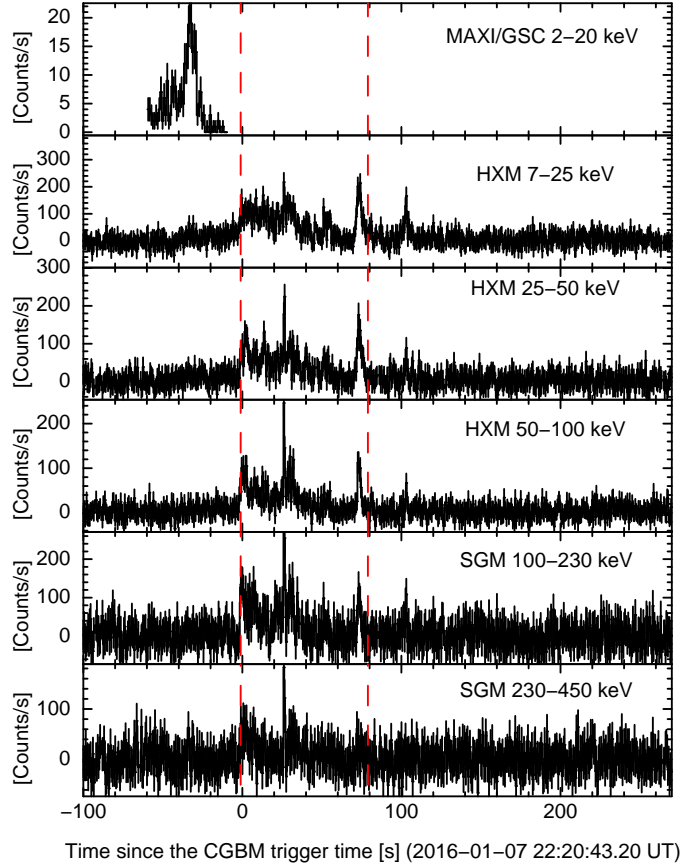}
 \end{center}
 \caption{The 1 s binning light curves of GRB 160107A which were observed by MAXI and CGBM.  
 Red dashed lines show the time interval of the time-averaged spectrum.  The MAXI light curve (top panel) is 
 not corrected for the time-variable effective area.  (Color online)}
\label{light_curve}
\end{figure*}

\begin{figure*}[h]
 \begin{center}
  \includegraphics[width=80mm,  angle=270]{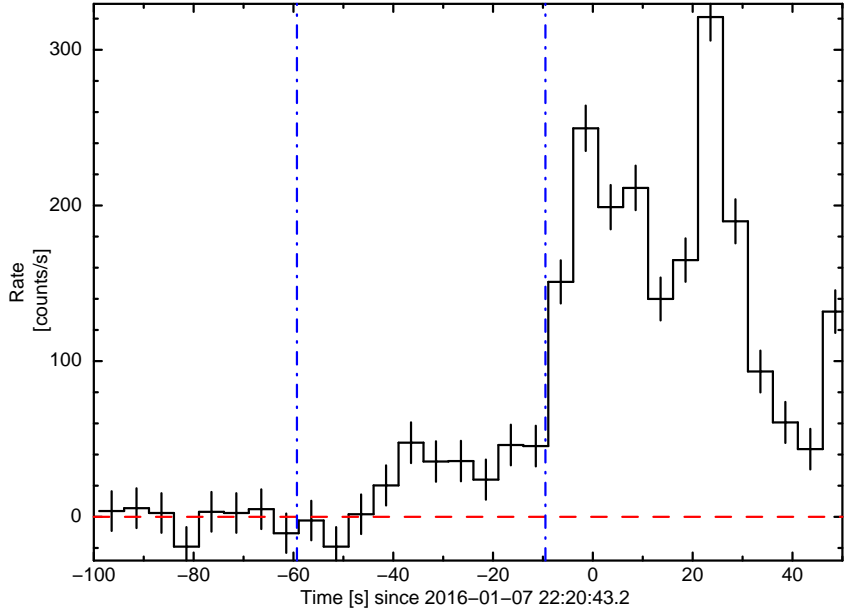}
 \end{center}
 \caption{The 5 s binned HXM light curve of GRB 160107A (7 -- 100 keV).  A red dashed line shows the background level.  
Blue dot-dashed lines represent the observation time-window of MAXI.  The prior emission observed by MAXI is visible in the 
HXM light curve.  (Color online)}
\label{hxm_lc_zoom}
\end{figure*}

GRB~160107A was detected by CGBM on 2016 January 7 at  22:20:43.20 (UT) \citep{Nakahira_2016}.  
The burst was also detected by MAXI/GSC \citep{Nakagawa_2016}, {\it Fermi}/GBM \citep{Veres_2016}, {\it INTEGRAL} (SPI-ACS) and {\it Wind}/KONUS \citep{Golenetskii_2016}.  Figure \ref{light_curve} shows the light curve of GRB~160107A 
based on the CGBM and MAXI/GSC data.  The CGBM light curve shows a complex structure with multiple 
overlapping pulses.  We use the {\it Swift} analysis tool, {\tt battblocks}, to measure the $T_{\rm 90}$ of the SGM data.  
The no-background subtracted light curve in 0.125 s binning over the 40--1000 keV band is used.  
The $T_{\rm 90}$ measured by the SGM is 73.5 $\pm$ 19.0 s (1 $\sigma$).  Thus, GRB~160107A 
is classified as a long GRB.  Interestingly, as seen in figure \ref{light_curve}, the emission seen in the MAXI/GSC 
was $\sim$45 s earlier than the CGBM trigger time.  At the time of the MAXI detection, a faint emission is visible 
in the HXM light curve (figure \ref{hxm_lc_zoom}).  The significance of this weak emission in the 
HXM is 5.5$\sigma$ in the 7 - 100 keV band.  Note that the field of view of the MAXI/GSC camera moved outside 
the location of GRB~160107A around 20 s before the CGBM trigger time.  Therefore, the main part of the emission was  
not detected by the MAXI/GSC.

\begin{figure*}[h]
 \begin{center}
  \includegraphics[width=100mm]{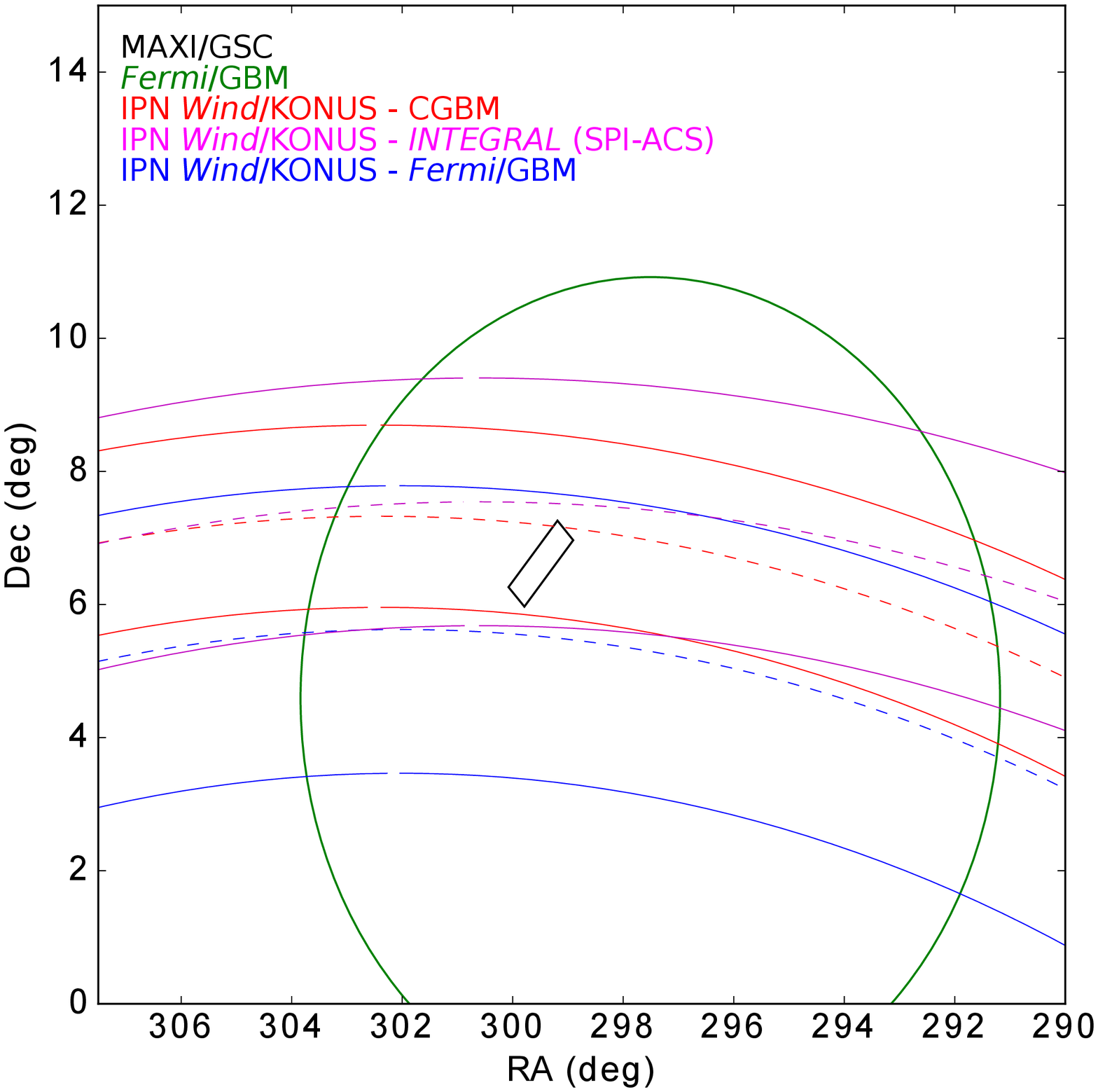}
 \end{center}
 \vspace{1.0cm}
 \caption{The localization of GRB 160107A.  Black and green line represent the MAXI/GSC error box and {\it Fermi}/GBM error circle.  
 Red, magenta and blue line show the IPN localizations of {\it Wind}/KONUS - CGBM,  {\it Wind}/KONUS - {\it INTEGRAL} (SPI-ACS) 
 and {\it Wind}/KONUS - {\it Fermi}/GBM, respectively.  (Color online)}
\label{position}
\end{figure*}

The best fit 90\% error region based on the MAXI/GSC data is described by a rectangular region 
with the following four corners: (R.A., Dec. ) = (298$^{\circ}.$905, 6$^{\circ}.$966), 
(299$^{\circ}.$190, 7$^{\circ}.$260), (300$^{\circ}$.076, 6$^{\circ}$.261) and 
(299$^{\circ}$.789, 5$^{\circ}$.968) (J2000.0).  The {\it Fermi}/GBM ground localization is 
(R.A., Dec.) = (297$^{\circ}$.510, +4$^{\circ}$.590) (J2000.0) with $2^{\circ}.6$ error radius 
(1 $\sigma$) plus systematic error of $\sim 3^{\circ}.7$ \citep{Connaughton_2015} which is consistent with the MAXI/GSC error region.  We further performed 
the localization analysis using the Interplanetary Network (IPN).  
We cross-correlated the light curve data of the ${\it Fermi}$/GBM and the ${\it Wind}$/Konus, and 
the SGM and the ${\it Wind}$/Konus, the {\it INTEGRAL} (SPI-ACS) and the ${\it Wind}$/Konus.  Figure \ref{position} shows the IPN position annuli overlaid with the MAXI/GSC and the ${\it Fermi}$/GBM best positions. The IPN location is 
consistent with the MAXI/GSC error region, but the MAXI/GSC error region provides 
the best localization of this GRB.  Based on the temporal and the positional coincidence, we concluded that 
the MAXI/GSC emission is clearly associated to GRB 160107A.  

\subsection{Spectrum}

First, we investigated the spectral properties of the main part of the emission using 
the CGBM data.  The spectral files extracting the data between $T_{0}-$1.1 s and $T_{0}+$78.9 s 
are created for HXM1, HXM2 and SGM, where $T_{0}$ is the trigger time of CGBM (January 7 at 22:20:43.20).  
The background files are generated for each spectral channel by 
fitting the channel separately with a second order polynomial function before the burst ($T_{0}-269.1$ s to 
$T_{0}-81.1$ s) and after the burst ($T_{0}+230.9$ s to $T_{0}+798.9$ s).  The gain-correction is applied to those spectral files by fitting the background 
lines at 1.47 MeV due to  $^{138}$La in the LaBr$_3$ (Ce) crystal for the HXM data \citep{Quarati_2012} and 2.2 MeV due to activation of the  BGO crystal 
for the SGM data.  We examined the gain-corrected spectra using the background lines at 35.5 keV and 511 keV, and found the energy of those lines 
to be accurate within 5\%.  The detector response matrices (DRMs) of CGBM are developed using a Monte-Carlo simulator based on the {\it GEANT4} software package \citep{agostinelli2003}.  Since the DRMs are sensitive to the incident angle of the event, we run the simulator using the best burst position from MAXI, (R.A., Dec.) = (299$^{\circ}$.670, 
6$^{\circ}$.413) (J2000.0).  The incident angles of GRB~160107A at the detector plane of HXM and SGM 
are ($\theta$, $\phi$) = (23$^{\circ}$.2, 174$^{\circ}$.6) and (13$^{\circ}$.3, 170$^{\circ}$.6) where 
$\theta$ and $\phi$ are zenith and azimuth angles.  The zenith direction of CALET corresponds to $\theta=0^{\circ}$.  
The front direction of CALET corresponds to $\phi=0^{\circ}$.  Since the variation in the incident angle $\theta$ during 
the burst interval ($\sim 80$ s) is less than $\sim 1^{\circ}.5$, we used the DRM for the fixed position.  Details about the 
CGBM DRMs, the calibration status and the results of the cross spectral calibrations using the simultaneously detected 
bright GRBs with the {\it Swift}/BAT data are described in appendices 1 and 2.  Since there is uncertainty due to absorption 
by structures around detectors in the low energy region,  we only used the data above 30 keV for HXM and above 100 keV 
for SGM.  XSPEC version 12.9.1 is used in the spectral analysis.  

Table \ref{tab:fit_cgbm} presents the results of the time-integrated spectral analysis of 
the burst.  The spectrum is best fitted by a power-law times an exponential cutoff (CPL) function.  The $\chi^2$ improvement 
in a CPL over a simple power-law (PL) fit is 8.42 in 1 degree of freedom.  
A Band function results in no better $\chi^2$ than CPL.  
We simulated 10,000 spectra inputting the best fit parameters of a CPL model using the same background spectrum 
generated in the analysis.  
We fit the simulated spectra by both PL and CPL models, and calculate the number of the simulated 
spectra which exceed $\Delta \chi^2 (=\chi^2 (\mathrm{CPL}) - \chi^2(\mathrm{PL}))$ of 8.42.  We found that 
4,459 simulated spectra exceed $\Delta \chi^2$ of 8.42.  This corresponds to the significance of the improvement of 44.6\%.  
Therefore, the improvement $\chi^2$ of a CPL over a PL in the observed spectrum is not highly significant based 
on the simulation study.  However, 
%
the best fit photon index $\alpha$ 
of $-1.7 ^{+0.3}_{-0.2} $ and $E_{\rm peak}$ of $102^{+48}_{-40}$ keV in a CPL model are consistent with {\it Wind}/KONUS and 
{\it Fermi}/GBM (\cite{Golenetskii_2016}; \cite{Gruber_2014}; \cite{Kienlin_2014}; \cite{Bhat_2016}), with $\alpha$ steeper 
than typical values of the low energy photon index for long GRBs \citep{kaneko2006}.  The fluence and the 1-second peak flux 
calculated in the best fit CPL are $1.6 \pm 0.1 \times 10^{-5}$ erg cm$^{-2}$ ($1\sigma$) and $1.6 \pm 0.1 \times 10^{-6}$ erg cm$^{-2}$ s$^{-1}$ 
($1\sigma$) at the 30 - 500 keV band.  The 1-second peak flux was measured between $T_{0}+26.2$ s and $T_{0}+27.2$ s.  

\begin{figure*}[h]
 \begin{center}
  \includegraphics[width=80mm, angle=270]{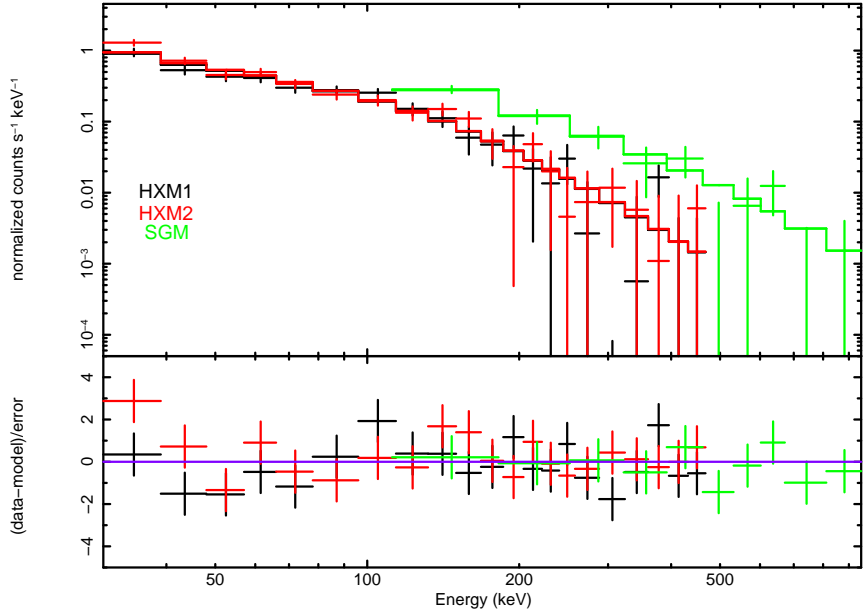}
 \end{center}
 \caption{The CGBM time-averaged spectrum of GRB 160107A.  Black, red and green points are the HXM1, HXM2 and 
 the SGM data, respectively.  The best fit model is a CPL.  (Color online)  }
 \label{spec_cgbm}
\end{figure*}

Next, we examined the MAXI/GSC spectrum at the time of the MAXI detection.  We used the MAXI standard software 
for creating the spectrum.  
The calibrated MAXI/GSC event data were downloaded from the Data Archive Transmission System (DARTS).\footnote{http://darts.isas.jaxa.jp/pub/maxi}  
The MAXI/GSC camera ID 4 and 5 detected the photons from the GRB.  After the observation information files were 
generated by {\tt mxscancur}, the exposure maps (WMAP files) for individual camera were created by {\tt mxgtiwmap}.  
The source and the background spectral files were extracted by {\tt xselect} from the calibrated MAXI/GSC event file.  
The source region is selected as a circle of a radius of 1$^{\circ}$.6 at (R.A., Dec.) = (299$^{\circ}$.670, 6$^{\circ}$.413) (J2000.0) as a center.  The background region is an annulus of an outer radius of 3$^{\circ}$ and an inner radius of 1$^{\circ}$.6 
at (R.A., Dec.) = (299$^{\circ}$.670, 6$^{\circ}$.413) (J2000.0) as a center.  There is $\sim$10 \% uncertainty in the exposure time 
due to the position uncertainty.  The source spectrum is extracted when the effective area to the source position is $>$0.05 cm$^{2}$ 
which corresponds from  $T_{0}-59.3$ s to $T_{0}-9.5$ s.  The detector response files were generated by {\tt mxgrmfgen}.  
The energy range used in the spectral analysis of the MAXI/GSC spectrum is from 2 keV to 30 keV. 
There is no issue of the dead-time in the MAXI/GSC data in the count rate level of GRB 170107A.  

Table \ref{tab:fit_maxi} shows the results of the spectral analysis of the MAXI prior emission.  We fit the spectrum 
by a power-law, a blackbody, a cutoff power-law, Band function and a power-law plus blackbody model 
with the interstellar absorption, $N_{\mathrm H}$, which is the model ``wabs'' in {\tt xspec}.  The best fit spectral model 
of the MAXI/GSC data is the highly absorbed power-law model with $N_{\mathrm H}$ = $3.7_{-2.6}^{+3.2} \times 10^{22}$ cm$^{-2}$.  
However, we believe this best fit model is unphysical because of the unrealistically large $N_{\mathrm H}$.  This GRB is 
located at the galactic latitude of $-12^{\circ}$ and the calculated Galactic $N_{\mathrm H}$ by \citet{dl1990} 
at this direction is $1.4 \times 10^{21}$ cm$^{-2}$.  Furthermore, according to \citet{willingale2013}, the mean observed 
$N_{\mathrm H}$ of X-ray afterglow spectra observed by {\it Swift} XRT is $2.1 \times 10^{21}$ cm$^{-2}$.  
Our derived $N_{\mathrm H}$ of $3.7 \times 10^{22}$ cm$^{-2}$ corresponds to $4.7 \times 10^{-3}$ percentile 
of the $N_{\mathrm H}$ distribution of \citet{willingale2013}.  Therefore, we concluded that the absorption-like-feature apparently 
seen at low energy in the MAXI/GSC spectrum is not due to the real absorption by the interstellar medium and 
the host galaxy of GRB 160107A.  
On the other hand, although a residual is seen at high energies, 
the spectrum can be also well fitted by a blackbody with the temperature of 1.2 keV.  It is possible to reduce this residual in 
the blackbody fit by adding a power-law component.  
No significant improvement in $\chi^2$ is seen by the Band function over a simple power-law fit.  

\begin{figure*}[h]
\begin{center}
\includegraphics[width=10cm,angle=0]{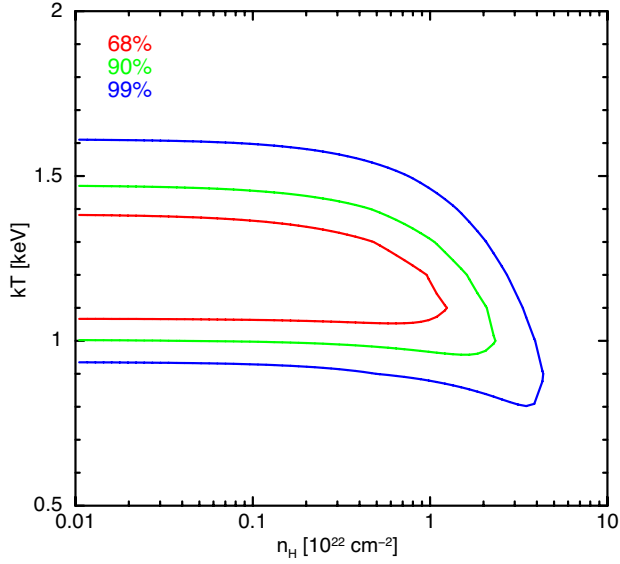}
\end{center}
\caption{Confidence contour between a blackbody temperature and $N_{\mathrm H}$ at the spectrum of the prior emission interval.} \label{kt_nh_conf}
\end{figure*}

We also performed joint fitting analysis including the HXM data.  The time interval of the HXM spectrum 
is between $T_{0}-61.0$ s and $T_{0}-13.1$ s.  The results of the joint fitting are shown in table \ref{tab:fit_maxi}.  
We found that the best fit spectral model is a blackbody plus a power-law model in the joint fit analysis.  
The photon index of $\alpha = -1.6 \pm 0.3$ and a temperature $kT = 1.0_{-0.2}^{+0.3}$ keV are consistent with 
the results from the MAXI/GSC data alone.  
The $\chi^2$ improvement of an absorbed blackbody plus a power-law model over an absorbed power-law model 
is 5.10 in 2 degrees of freedom.  
We simulated 10,000 spectra inputting the best fit parameters of an absorbed blackbody plus a power-law model.  
We fit the simulated spectra by both an absorbed blackbody plus a power-law model and an absorbed power-law 
model, and calculate the number of the simulated spectra which exceed $\Delta \chi^2$ of 5.10.  We found that 
3,056 simulated spectra exceeding $\Delta \chi^2$ of 5.10.  The significance of the improvement is 30.6\%.  
Addition to the discussion regarding $N_{\mathrm H}$ in the previous paragraph, the best representative model of 
the prior emission spectrum 
is a blackbody plus a power-law model, with the thermal emission 
superimposed on the non-thermal component at $\sim$45 seconds before the main burst episode.  
As shown in Figure \ref{kt_nh_conf}, our derived blackbody temperature is not sensitive to the value 
of $N_{\mathrm H}$.  

\begin{figure*}[h]
 \begin{center}
  \includegraphics[width=80mm,angle=270]{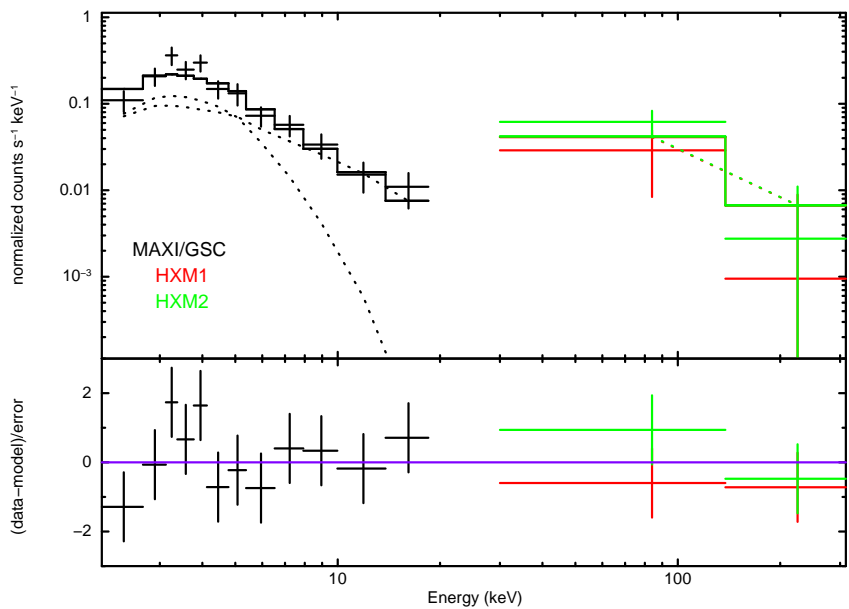}
 \end{center}
 \caption{The time-averaged spectrum of the prior emission with a best-fit PL+BB model.  Black, red and green points are the MAXI/GSC, the HXM1 and the HXM2 data, 
 respectively.  The best fit model is PL + BB. (Color online) }\label{spec_maxi}
\end{figure*}

\begin{table*}[htbp]
\tbl{Time-averaged spectral parameters of the CGBM spectrum.}{
 \begin{tabular}{ccccccc}
   \hline
  Model & $\alpha$ & $\beta$ & $\ep$ & A\footnotemark[$*$] & SGM constant factor &$\chi ^{2}/$ d.o.f. \\
              &               &              & [keV]   & [ph cm$^{-2}$ s$^{-1}$ keV$^{-1}$] & \\
  \hline
 PL\footnotemark[$\dagger$]  & $-2.1 \pm 0.1$&$\singlebond$&$\singlebond$&67$^{+39}_{-24}$&1.33$^{+0.31}_{-0.27}$&54.04 / 49\\
 CPL\footnotemark[$\ddagger$]  &$-1.7^{+0.3}_{-0.2}$&$\singlebond$&102$^{+48}_{-40}$&19$^{+28}_{-12}$&1.42$^{+0.31}_{-0.27}$&45.62 / 48\\
 BAND\footnotemark[$\S$]  &$-1.8 ^{+0.2}_{-0.3}$& $-2.3$ (fixed) &$94^{+74}_{-83}$&6.0$^{+2.8}_{-1.0}$ $\times 10^{-3}$ &1.35$^{+0.29}_{-0.26}$&48.92 / 48\\
     \hline
  \end{tabular}}\label{tab:fit_cgbm}
\begin{tabnote}
\footnotemark[$*$] The normalization of a power-law and a cutoff power-law model is calculated at 1 keV.  Whereas, the 
normalization of a Band function is calculated at 100 keV.  \\
\footnotemark[$\dagger$] A power-law model.\\
\footnotemark[$\ddagger$] A power-law times exponential cutoff model.\\
\footnotemark[$\S$] Band function.\\
\end{tabnote}
\end{table*}

\begin{table*}[htbp]
\tbl{Fit parameters for the MAXI prior emission spectrum.}{
 \begin{tabular}{ccccccccccc}
   \hline
  Instruments&Model & nH & $\alpha$ &$\beta$& $\ep$ & A$_{\rm PL}$\footnotemark[$*$]  & kT & A$_{\rm BB}$\footnotemark[$\dagger$]  & $\chi ^{2}/$ d.o.f. \\
            & & [$10^{22}$ cm$^{-2}$ ] && & [keV] & [ph cm$^{-2}$ s$^{-1}$ keV$^{-1}$] & [keV] & &\\
  \hline
   MAXI&PL\footnotemark[$\S$] &3.7 $^{+3.2}_{-2.6}$&$-2.7^{+0.6}_{-0.8}$&$\singlebond$&$\singlebond$&12$^{+35}_{-8}$&$\singlebond$&$\singlebond$&9.18 / 9\\
  MAXI&PL &0.14 (fixed)&$-2.0 \pm 0.2$&$\singlebond$&$\singlebond$&2.8$^{+1.2}_{-0.9}$&$\singlebond$&$\singlebond$&14.64 / 10\\
  MAXI&BB\footnotemark[$\|$] &0.14 (fixed)&$\singlebond$&$\singlebond$&$\singlebond$&$\singlebond$&$1.2 \pm 0.2$&0.11$\pm 0.01$&21.21 / 10\\ 
  MAXI&CPL\footnotemark[$\#$] &0.14 (fixed)&$-1.2_{-0.9}^{+1.4}$&$\singlebond$&$>2$&1.9$^{+1.6}_{-1.0}$&$\singlebond$&$\singlebond$&13.03 / 9\\
  MAXI&BAND\footnotemark[$**$] &0.14 (fixed)&$+1.8^{+0.6}_{-2.5}$&$-2.4^{+0.4}_{-0.6}$&$<5$&$<124$&$\singlebond$&$\singlebond$&8.49/8\\ 
 MAXI&PL+BB &0.14 (fixed)&$-0.6_{-1.2}^{+3.5}$&$\singlebond$&$\singlebond$&0.09$^{+1.67}_{-0.09}$&1.0$\pm 0.2$&0.08$^{+0.03}_{-0.05}$&8.45 / 8\\
  MAXI&PL+BB &0.14 (fixed)&$-1.0$ (fixed)&$\singlebond$&$\singlebond$&0.2$\pm 0.1$&0.9$\pm 0.2$&0.07$\pm 0.02$&8.57 / 9\\
  MAXI \& HXM&PL &0.14 (fixed)&$-1.9 \pm 0.2$&$\singlebond$&$\singlebond$&2.6$^{+1.0}_{-0.7}$&$\singlebond$&$\singlebond$&16.83 / 14\\
  MAXI \& HXM&BB &0.14 (fixed)&$\singlebond$&$\singlebond$&$\singlebond$&$\singlebond$&1.2$\pm 0.2$&0.11$\pm 0.01$&31.83 / 14\\ 
  MAXI \& HXM&CPL &0.14 (fixed)&$-1.88$\footnotemark[$\ddagger$] &$\singlebond$&not determined&2.6\footnotemark[$\ddagger$]&$\singlebond$&$\singlebond$&16.84 / 13\\ 
  MAXI \& HXM&BAND&0.14 (fixed)&$+1.7^{+0.6}_{-3.2}$&$-2.1^{+0.2}_{-0.4}$&$<6$&$56^{+51}_{-56}$&$\singlebond$&$\singlebond$&13.47/12\\ 
  MAXI \& HXM&PL+BB &0.14 (fixed)&$-1.6 \pm 0.3$&$\singlebond$&$\singlebond$&0.9$^{+1.2}_{-0.6}$&1.0$^{+0.3}_{-0.2}$&0.05$^{+0.03}_{-0.04}$&11.73 / 12\\
  \hline
  \end{tabular}}  \label{tab:fit_maxi}
\vspace{0.5cm}
\begin{tabnote}
\footnotemark[$*$] A normalization of either a power-law or a power-law times exponential cutoff model at 1 keV.  Whereas, the 
normalization of a Band function is calculated at 10 keV. \\
\footnotemark[$\dagger$] A normalization of a blackbody in the unit of  $L_{39} / D_{10}^{2}$ where 
$L_{39}$ is the source luminosity in units of $10^{39}$ erg s$^{-1}$ and $D_{10}$ is the distance of the source in 
units of 10 kpc\\
\footnotemark[$\ddagger$] An error is not available.\\
\footnotemark[$\S$] A power-law model.\\
\footnotemark[$\|$] A blackbody.\\
\footnotemark[$\#$] A power-law times exponential cutoff model.\\
\footnotemark[$**$] Band function.\\
\end{tabnote}
\end{table*}

\section{Discussion}
The MAXI/GSC detection prior to the main burst episode is achieved in coordination with the high sensitity 
soft X-ray survey data of the MAXI/GSC and the wide field hard X-ray survey data of the CGBM, both of which 
are physically located on the same platform.  The CGBM provides the data from 7 keV up to 
20 MeV for a bright transient source  whereas MAXI/GSC data provides 
simultaneous data from 2 keV up to 30 keV.  The field of view of the 
MAXI/GSC cameras always overlap with that of the CGBM.  Therefore, we would expect to have GRBs observed 
simultaneously by CGBM and MAXI/GSC.  We estimate the number of simultaneously observed GRBs  as follows: 
The instantaneous field of views of the SGM and the MAXI/GSC are 20000 deg$^{2}$ ($\sim$2 $\pi$) and 
480 deg$^{2}$ ($1^{\circ}.5 \times 160^{\circ}$ in two directions).  
The observation efficiencies are 40\% for the MAXI/GSC \citep{Sugizaki_2011} and 60\% for the CGBM.  Since the CGBM detects $\sim$50 GRBs per year, we expect $\sim$1 GRB per year to be detected simultaneously.  However, as we show here for GRB 160107A, the timing is not necessary matched between two instruments.  Furthermore, to confirm this fact, the MAXI/GSC also detected the emission at 320 s after the main burst episode for GRB 160509A \citep{ono2016}.  So far, we have four simultaneously detected GRBs between the CGBM and the MAXI/GSC in one year of the overlapping operation.  

Following the argument of \citet{peer2007}, we estimate the photospheric radius 
from the observed parameters of the blackbody component.  The photospheric radius 
$R_{\mathrm{ph}}$ can be estimated by the following equation:  
$R_{\mathrm{ph}} = \gamma_{0} \, d_{\mathrm{L}} \, \mathcal{R} / \xi \, (1+z)^2$ 
where $\gamma_{0}$ is a bulk Lorentz factor, $d_{\mathrm{L}}$ is a luminosity distance, 
$\xi$ is a geometric factor of order of unity (we assume $\xi = 1$), $z$ is a redshift and 
$\mathcal{R}$ ($\equiv (F_{\mathrm{BB}}/\sigma \, T_{\mathrm{BB}}^4)^{0.5}$) is the ratio 
between the blackbody flux $F_{\mathrm{BB}}$ and 
$\sigma \, T_{\mathrm{BB}}^4$ where $\sigma$ is the Stefan-Boltzmann constant and 
$T_{\mathrm{BB}}$ is the blackbody temperature.  From the observed blackbody component, 
we derive $\mathcal{R} = 2.9 \times 10^{-17}$.  Assuming a bulk Lorentz factor of 100 and 
a redshift of 1, together with the cosmological parameters of $\Omega_{m} = 0.27$, 
$\Omega_{\Lambda} = 0.73$ and $H_{0} = 71$ km s$^{-1}$ Mpc$^{-1}$, the photospheric radius is calculated as 
\begin{equation}
R_{\mathrm{ph}} = 1.5 \times 10^{13} \, \mathrm{cm}\, \left(\frac{\gamma_{0}}{100}\right) \, \left(\frac{d_{\mathrm{L}}}{2.0 \times 10^{28} \, \mathrm{cm}}\right) \left(\frac{1+z}{2}\right)^{-2} \, \left(\frac{\mathcal{R}}{2.9 \times 10^{-17}}\right).  
 \end{equation}
The radius of the main burst emission site is estimated by the 45 s delay which we 
observed between the MAXI/GSC and the CGBM data.  The difference of 
the arrival time of the photons emitted between $R_{0}$ and $aR_{0}$ ($a > 1$) to the 
observers can be expressed as the radial time scale \citep{piran1999}, $\Delta T = 
[(a^{4} - 1)/4](R_{0}/2\gamma_{0}^{2} c)$ where $\gamma_{0}$ is a bulk Lorentz 
factor at $R_{0}$ and $c$ is a speed of light.  Substituting $\Delta T$ as the observed 
delay time $\Delta T_{\mathrm{obs}}$ of 45 s ($\Delta T = \Delta T_{\mathrm{obs}} / (1+z)$) 
and $R_{0}$ as the derived photospheric radius of $1.5 \times 10^{13}$ cm, 
the parameter $a$ is estimated as 
\begin{equation}
a \sim 8 \, \left( \frac{\Delta T_{\mathrm{obs}}}{45 \, {\mathrm s}}  \right)^{\frac{1}{4}} \left( \frac{1+z}{2} \right)^{- \frac{1}{4}} \left( \frac{\gamma_{0}}{100} \right)^{\frac{1}{2}} \left( \frac{R_{\mathrm{ph}}}{1.5 \times 10^{13} \, {\rm cm}} \right)^{- \frac{1}{4}}.
\end{equation}
Since the radius to the main burst episode site is $aR_{\mathrm{ph}}$, its radius is estimated to be $\sim1.2 \times 10^{14}$ cm.  
This estimated radius is consistent with the radius between internal shocks in the case of $\gamma_{0} = 100$ \citep{piran1999}.  
Note, however, that our discussion relies on the assumption of a redshift.  Unfortunately, the sky location of GRB 160107A 
at the time of the trigger was too close to the Sun to perform the follow-up observations from the ground facilities.  
As shown here for GRB 160107A, the soft X-ray data of MAXI can provide an important clue in understanding 
the prompt emission from GRBs.  Therefore, we strongly encourage the follow-up to MAXI detected GRBs to identify 
an afterglow and a host galaxy in order to have a secure redshift measurement.  

A possible origin of a thermal component is an emission from a mildly relativistic cocoon around a jet 
\citep{ramirez2002,nakar2017}.  According to the calculation of \citet{lazzati2017}, X-ray flux expected from 
a cocoon is $\sim10^{-14} - 10^{-12}$ erg cm$^{-2}$ s$^{-1}$ at a typical distance of 1 Gpc for long GRBs.  
Our detected thermal X-ray emission is $8 \times 10^{-9}$ erg cm$^{-2}$ s$^{-1}$ which is two order of magnitude 
brighter than the calculation.  Another interesting possibility is a supernova shock breakout emission.  The 
onset X-ray emission of a shock breakout is detected by {\it Swift} XRT for SN 2008D \citep{soderberg2008}.  
The peak X-ray luminosity of this shock breakout is $6 \times 10^{43}$ erg s$^{-1}$ which corresponds 
to $5 \times 10^{-13}$ erg cm$^{-2}$ s$^{-1}$ at 1 Gpc.  This estimated flux is three order of magnitude 
weaker than the thermal emission of GRB 160107A.  Therefore, either a cocoon or a shock breakout emission 
is expected to be below the detection sensitivity of MAXI/GSC at the typical distance of long GRBs, and 
is difficult to be the origin of the thermal emission of GRB 160107A.  

The blackbody temperature of 1 keV which we see in the GRB 160107A spectrum is one or two order 
of magnitudes smaller than the reported blackbody temperature in the BATSE (e.g., \cite{ryde2005}) and 
the Fermi-GBM (e.g., \cite{ryde2010,guiriec2013}) GRB spectra.  Our temperature is more 
consistent with the identified blackbody components reported by {\it HETE-2} \citep{shirasaki2008} 
and {\it Swift} \citep{starling2012}.  However, our understanding of the thermal emission 
in the prompt GRB spectrum is limited by a small number of incomplete (e.g, no redshift measurement) 
samples.  As we demonstrated here for GRB 160107A, the coordination of the scientific instruments on-board ISS 
becomes important to enhance the science outcome.  Joint observation by on-going X-ray missions 
like MAXI and NICER, and a future mission like TAO-ISS along with the hard X-ray survey instrument like 
CGBM will be crucial to collect high quality broad-band data related to the GRB emission.  

We would like to thank the anonymous referee for comments and suggestions that materially improved the paper.  
We would also like to thank D. Svinkin for providing the analysis of the IPN localization, and also useful comments and discussions. 
We gratefully acknowledge JAXA's contributions for CALET development and on-orbit operations. We express our sincere thanks 
to all of the CALET members for allowing us to use the CGBM data.  The CALET data used in this analysis are provided by the 
Waseda CALET Operation Center (WCOC) located at the Waseda University.  This work is supported by MEXT KAKENHI Grant 
Numbers 17H06357, 17H06362 (T.S.), and 24684015 (K.Y.).  This research was also supported by a grant from the Hayakawa Satio Fund awarded by the Astronomical Society of Japan (Y.K.).
\appendix
\section{Energy response function of CGBM}

The CGBM energy response function has been built by the simulator based on the {\it GEANT4} simulation package 
\citep{agostinelli2003}.  All the materials of the flight CGBM detectors are modeled in the simulator.  The entire CALET structure 
is included in the simulator, although, a simplified structure is used for the CAL (only precisely modeling the heavy elements which are 
responsible for the absorption and the scattering) in order to reduce the calculation speed \citep{Yamada_2017}.  
The detailed instrumental characteristics of HXM and SGM, based on the results of the pre-flight ground testing 
\citep{nakahira2017}, are also included in the simulator.  

\begin{figure*}[htbp]
   \begin{center}
   \includegraphics[width=100mm]{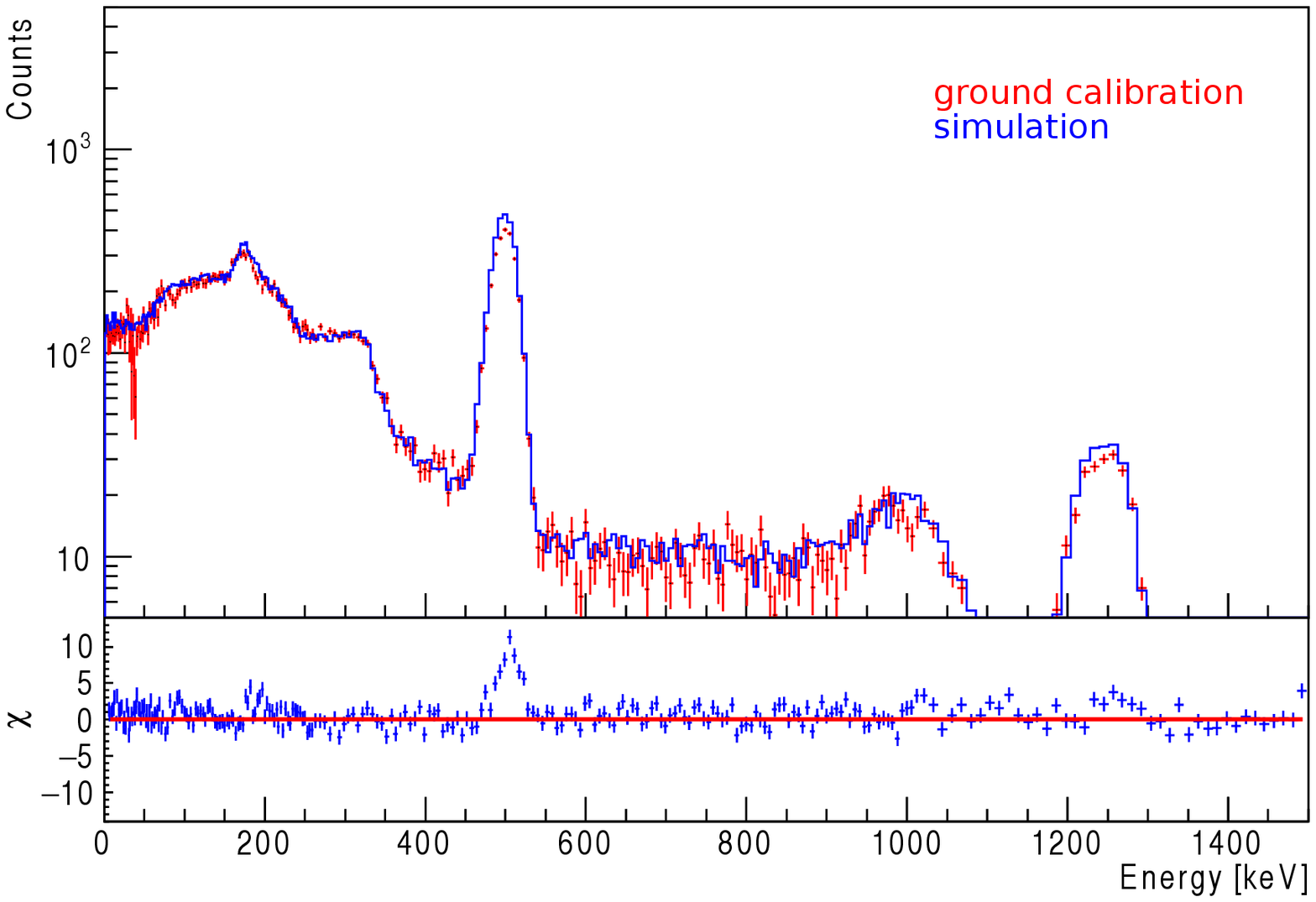}
   \\
   \includegraphics[width=100mm]{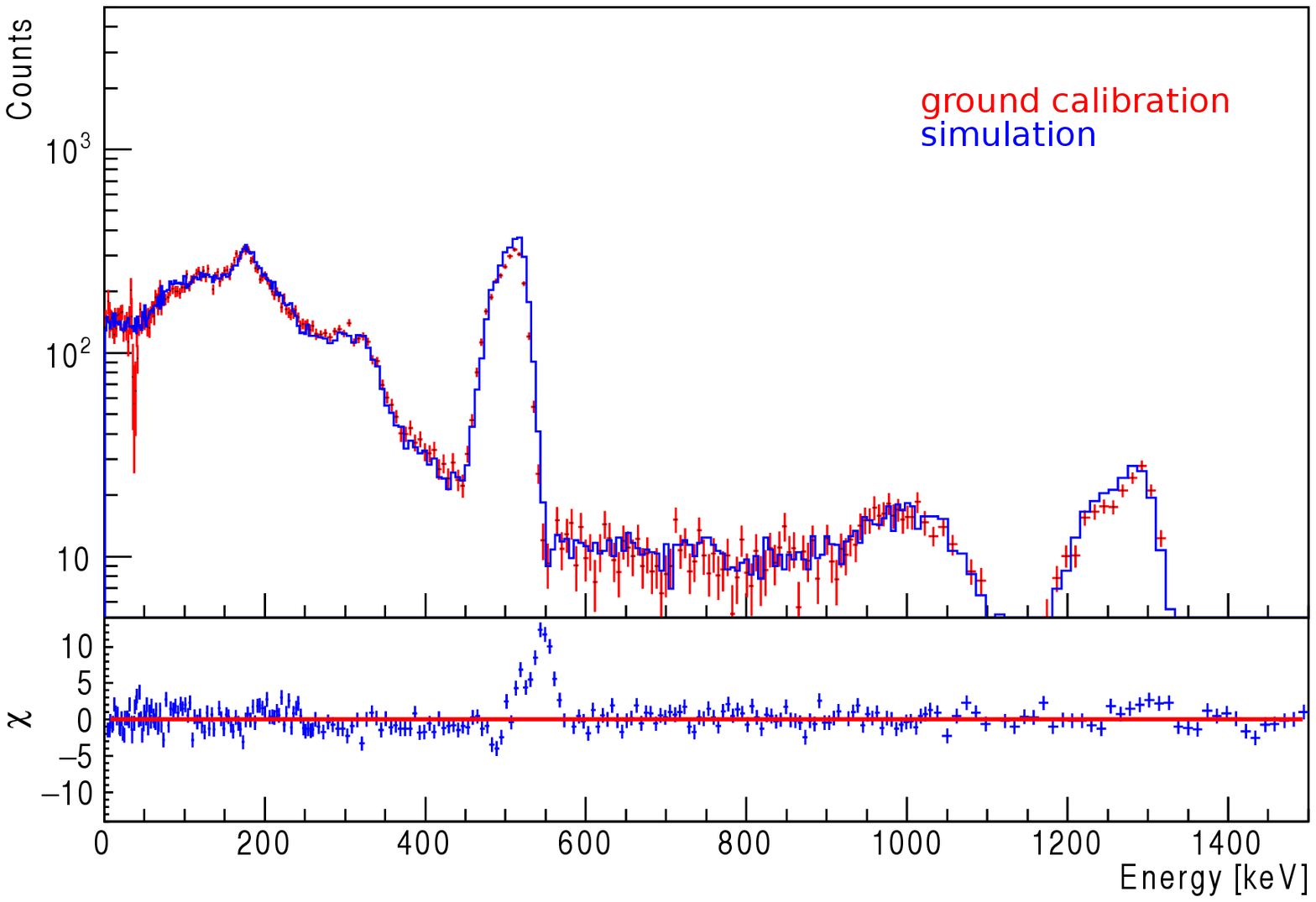}
   \\
   \includegraphics[width=100mm]{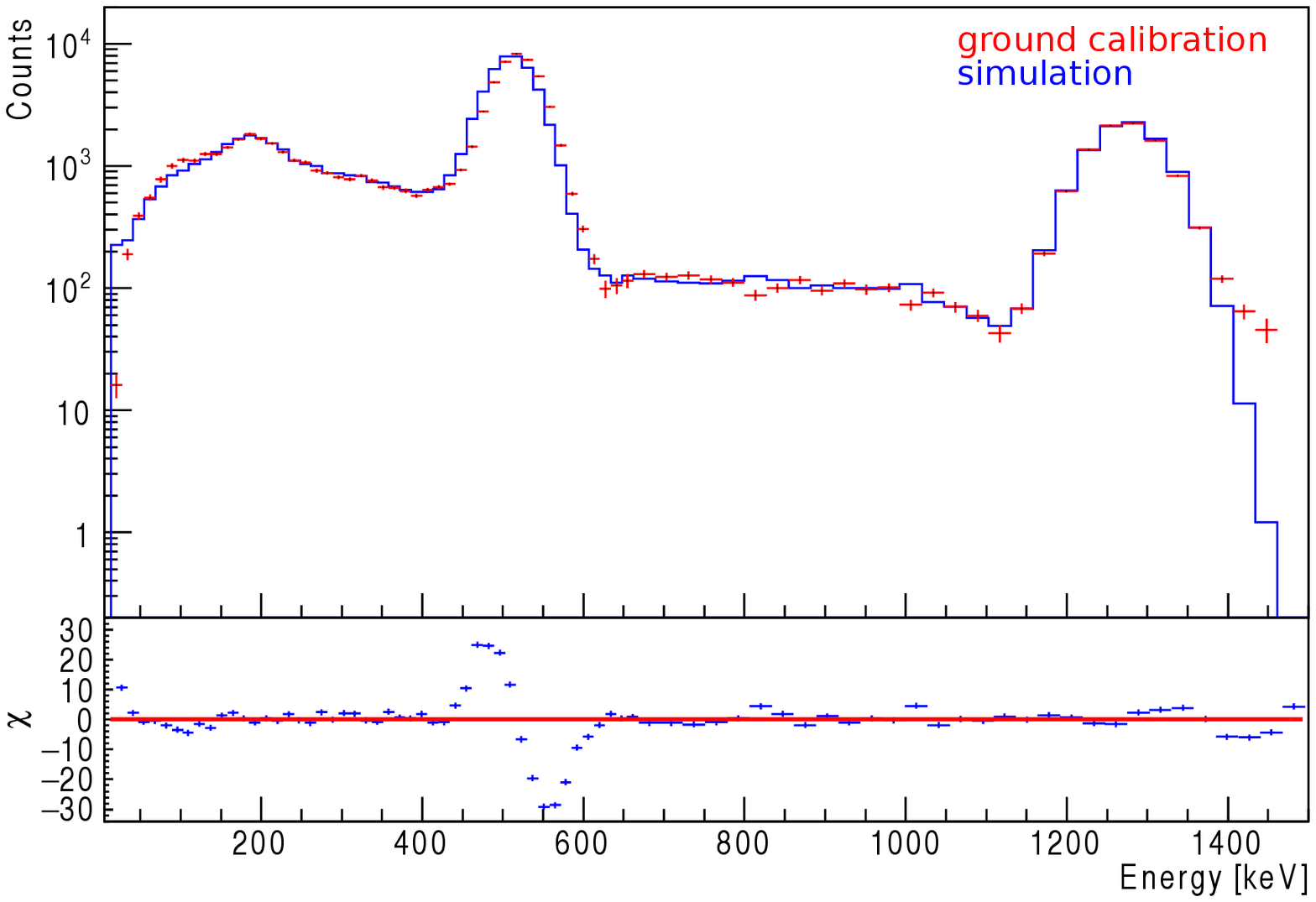}
 \end{center}
  \caption{The ground calibration (red) and the simulated (blue) spectra of $^{22}$Na. Top, middle, and bottom panels show spectrum of HXM1, HXM2 and SGM, 
  respectively. (Color online)}  
    \label{Na_spec}
\end{figure*}

Ground testing data were collected by irradiating the flight instruments with radioactive 
sources and soft X-rays from an X-ray generator.  Figure \ref{Na_spec} shows 
the comparison between the data and the simulated $^{22}$Na spectra for HXM1, HXM2 and SGM using the 
CGBM simulator.  As can be seen in the figure, both the energies and the normalizations of the 511 keV and the 
1275 keV lines agree quite well.  The difference in the line shape of the 511 keV line between HXM1 and HXM2 
is also well modeled in the simulator.  Despite a clear residual in the 511 keV line in the SGM, the overall shape 
of the spectrum shows a good agreement.  The net counts in the 511~keV peak in the simulated spectra 
are 20 \% and 3 \% larger than the calibration data of HXM and SGM.  The energy resolution of the 
simulated spectra agrees with the measured value to within 5\% for all detectors.  Therefore, the CGBM simulator 
has been demonstrated to reproduce the ground testing data accurately.  

\section{Spectral cross-calibration between CGBM and {\it Swift}/BAT using the simultaneously detected bright GRBs}
We use the bright GRBs simultaneously detected by the {\it Swift}/BAT since the systematic error in the energy response 
function has been studied in detail among the {\it Wind}/KONUS and the {\it Suzaku}/WAM using the GRB data \citep{sakamoto2011}.  
We identified two GRBs, GRB 161218A and GRB 170330A, for this investigation.  The incident angles to the HXM and the 
SGM boresights are 37$^{\circ}$ and 30$^{\circ}$ for GRB 161218A, and 32$^{\circ}$ and 35$^{\circ}$ for GRB 170330A.  
We extracted the time-averaged spectra for both the CGBM and the BAT using the exactly the same time interval.  The BAT 
spectrum is extracted from the event data using the standard procedure.\footnote{https://swift.gsfc.nasa.gov/analysis/threads/batspectrumthread.html}
As was done in the spectral analysis of GRB 160107A, the XSPEC (v12.9.1) is used in this joint spectral analysis.  

\begin{figure*}[htbp]
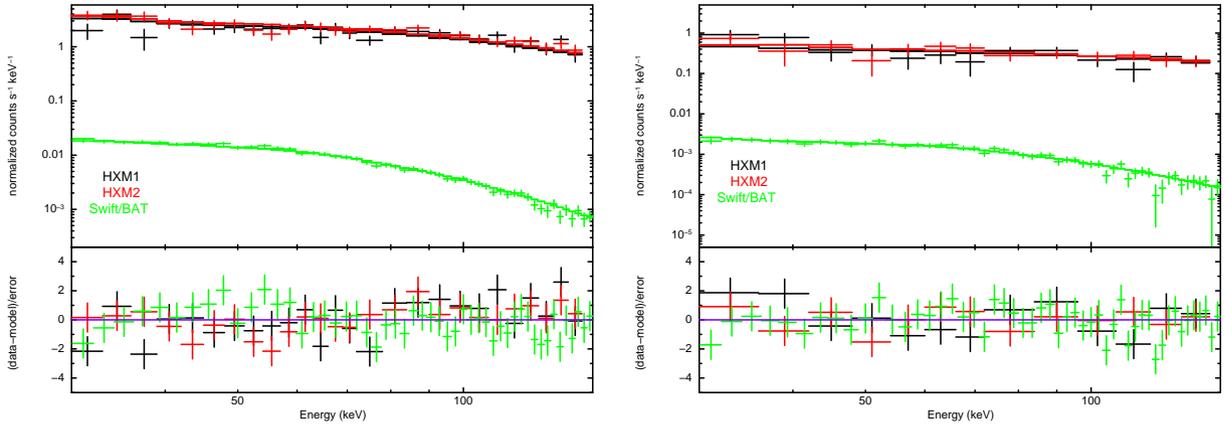

\begin{center}
  \includegraphics[width=55mm,angle=270]{./GRB161218A_nerr_low_only_HXM_BAT_30_150keV_cutplep_fit_paper.eps}
  \includegraphics[width=55mm,angle=270]{./GRB170330A_nerr_low_only_HXM_BAT_30_150keV_powerlaw_fit_paper.eps}
\end{center}
  \caption{Joint CGBM-{\it Swift}/BAT fitted spectra of GRB 161218A and GRB 170330A in the 30 -- 150 keV band.  
  Left panel shows the GRB 161218A spectrum with a best-fit CPL model.  Right panel shows the GRB 170330A spectrum and a best-fit PL model.  
  Black, red and green points are the HXM1, the HXM2 and the {\it Swift}/BAT data, respectively.  (Color online)}
    \label{GRB161218A_30_150keV}
\end{figure*}

First, we performed the joint spectral analysis only using the HXM data to focus on the investigation of the low energy 
response of the HXM.  The HXM spectrum is fitted between 30 and 150 keV.  Figure \ref{GRB161218A_30_150keV} 
shows the joint fit spectra and table \ref{tab:fit_grb161218a_grb170330a_hxm} summarizes the best fit parameters.  
Both the photon indices and the normalizations measured by the HXM are consistent with those of measured by {\it Swift}/BAT.  
Also, no significant residual is seen in both joint fitted spectra of GRB 161218A and GRB 170330A.  Therefore, we concluded 
that the systematic error in the low energy part of the HXM data (30--150 keV) is 10--20\% as is discussed in 
\citet{sakamoto2011} as the systematic uncertainty in the {\it Swift} BAT data.  

\begin{figure*}[htbp]
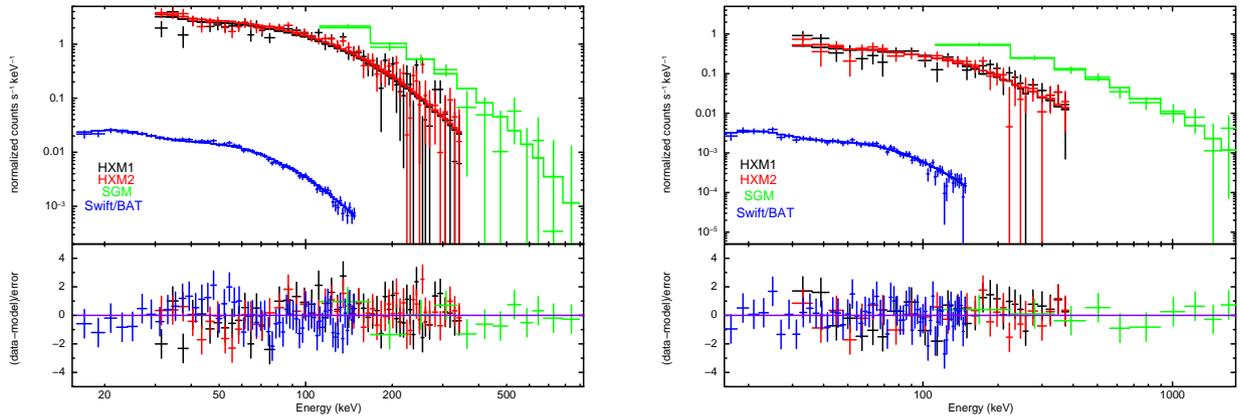

 \begin{minipage}{0.5\hsize}
  \begin{center}
 \includegraphics[angle=270,width=80mm]{./GRB161218A_nerr_low_only_cutplep_fit_paper.eps}
  \end{center}
 \end{minipage}
 \begin{minipage}{0.5\hsize}
  \begin{center}
   \includegraphics[angle=270,width=80mm]{./GRB170330A_nerr_low_only_cutplep_fit_paper.eps}
  \end{center}
  \end{minipage}
 \vspace{0.5cm}
  \caption{Joint CGBM-{\it Swift}/BAT fitted spectra of GRB 161218A and GRB 170330A including the SGM data.  
  Left panel shows the GRB 161218A spectrum with a CPL model.  Right panel shows the GRB 170330A spectrum 
  with a CPL model.  Black, red, green and blue points are the HXM1, the HXM2, the SGM and {\it Swift}/BAT data. (Color online)}
\label{GRB161218A_GRB170330A_joint}
\end{figure*}

Second, we included the SGM data in the fit.  In this three-instrument joint fit, we multiplied the model by a constant factor.  
We fixed the constant factor of BAT to unity.  Figure \ref{GRB161218A_GRB170330A_joint} shows the joint fit spectra and 
table \ref{tab:fit_grb161218a_grb170330a} summarizes the joint fits by different spectral models.  In the joint fitted spectra 
of GRB 161218A and GRB 170330A, the constant factors of HXM1 and HXM2 were consistent with unity, as is the case 
without the SGM data.  The centroids of the SGM constant factors of GRB161218A and GRB 170330A were 1.3 and 1.8,   
probably due to the incomplete modeling of the CALET in the simulator.  Furthermore, the time-dependent background 
variation due to charged particles makes it difficult to model the background spectrum accurately.  Since we are still working 
on understanding the influence of the structures around the detector and the uncertainty in the background modeling, we 
decided to multiply the model by a constant factor as a free parameter in the GRB 160107A data to take into account the 
currently unknown systematic uncertainty in the SGM data.  

\begin{table*}[htbp]
\tbl{Fit parameters of GRB 161218A and GRB 170330A with HXM and {\it Swift}/BAT.}{
 \scalebox{0.9}{
 \begin{tabular}{cccccccccc}
   \hline
  GRB name&instrument&models&C(HXM1)\footnotemark[$*$] &C(HXM2)\footnotemark[$\dagger$] &$\alpha$&$E_{\rm peak}$&norm& $\chi ^{2}/$ degree of freedom \\
   &             &&&               &              & [keV]   & [ph cm$^{-2}$ s$^{-1}$ keV$^{-1}$] & \\
  \hline
 GRB161218A&CGBM&CPL\footnotemark[$\ddagger$] &$\singlebond$&$\singlebond$&$-0.95^{+0.63}_{-0.33}$&not determined & 3.6$^{+10.0}_{-2.8}$ &44.75/49\\
                      & {\it Swift}/BAT&CPL&$\singlebond$&$\singlebond$&$-0.67 \pm 0.24$&127$^{+28}_{-15}$&2.1$^{+2.4}_{-1.1}$&33.95/48\\
                       & joint &CPL&1.01$^{+0.07}_{-0.06}$&0.98$\pm 0.07$&$-0.98 \pm 0.23$&193$^{+162}_{-45}$&5.4$^{+5.8}_{-2.8}$&106.15/98\\\hline
 GRB170330A&CGBM&PL\footnotemark[$\S$] &$\singlebond$&$\singlebond$&$-1.22 \pm 0.25$&$\singlebond$&1.5$^{+3.1}_{-1.1}$&22.97/24\\
                       & {\it Swift}/BAT&PL&$\singlebond$&$\singlebond$&$-1.17 \pm 0.09$&$\singlebond$&1.16$^{+0.50}_{-0.35}$&42.11/49\\
                       & joint &PL&0.95$\pm 0.14$&1.04$\pm 0.15$&$-1.17 \pm 0.08$&$\singlebond$&1.18$^{+0.48}_{-0.34}$&64.52/73\\
     \hline
  \end{tabular}} \label{tab:fit_grb161218a_grb170330a_hxm}
  }
  \begin{tabnote}
\footnotemark[$*$] A constant factor of HXM1.\\
\footnotemark[$\dagger$]  A constant factor of HXM2.\\
\footnotemark[$\ddagger$]  A power-law times exponential cutoff model.\\
\footnotemark[$\S$]  A power-law model.\\
\end{tabnote}
\end{table*}

\begin{table*}[htbp]
\tbl{Fit parameters of GRB 161218A and GRB 170330A from joint fit with HXM, SGM and {\it Swift}/BAT}{
 \scalebox{0.85}{
 \begin{tabular}{cccccccccc}
   \hline
  GRB name&models&C(HXM1)\footnotemark[$*$]&C(HXM2)\footnotemark[$\dagger$]&C(SGM)\footnotemark[$\ddagger$]&$\alpha$&$\beta$&$E_{\rm peak}$&norm& $\chi ^{2}/$ degree of freedom \\
& &&&                &               &              & [keV]   & [ph cm$^{-2}$ s$^{-1}$ keV$^{-1}$] & \\
  \hline
 GRB161218A&PL\footnotemark[$\S$]&$0.87 \pm 0.06$&$0.93 \pm 0.06$&0.37$\pm 0.06$&$-1.36 \pm 0.03$&$\singlebond$&$\singlebond$&$17 \pm 2$&556.90/171\\
                       &CPL\footnotemark[$|$]&$1.03 \pm 0.06$&$1.09 \pm 0.06$&$1.25^{+0.14}_{-0.13}$&$-0.69^{+0.09}_{-0.08}$&$\singlebond$&$148^{+13}_{-11}$&$2.1^{+0.6}_{-0.5}$&164.92/170\\
                       &BAND\footnotemark[$\#$]&$1.01 \pm 0.06$&$1.08 \pm 0.06$&$1.10 \pm 0.12$&$-0.62^{+0.11}_{-0.12}$&$-2.3$ (fixed)&$136^{+14}_{-12}$&$0.10^{+0.02}_{-0.01}$&194.67/170\\\hline
  GRB170330A&PL&$1.06^{+0.14}_{-0.13}$&$1.10^{+0.15}_{-0.14}$&$0.99^{+0.20}_{-0.18}$&$-1.25 \pm 0.05$&$\singlebond$&$\singlebond$&$1.6^{+0.4}_{-0.3}$&169.79/119\\
                       &CPL&$1.09^{+0.14}_{-0.13}$&$1.14 \pm 0.14$&$1.82^{+0.29}_{-0.26}$&$-0.98 \pm 0.08$&$\singlebond$&$463^{+123}_{-87}$&$0.63^{+0.22}_{-0.17}$&93.03/118\\
                       &BAND&$1.09^{+0.14}_{-0.13}$&$1.14 \pm 0.14$&$1.84^{+0.31}_{-0.27}$&$-0.97^{+0.10}_{-0.09}$&$-2.3$ (fixed)&$423^{+129}_{-95}$&6.9$^{+0.8}_{-0.6} \times 10^{-3}$&93.07/118\\
     \hline
  \end{tabular}} \label{tab:fit_grb161218a_grb170330a}
  }
  \begin{tabnote}
\footnotemark[$*$] A constant factor of HXM1.\\
\footnotemark[$\dagger$] A constant factor of HXM2.\\
\footnotemark[$\ddagger$] A constant factor of SGM.\\
\footnotemark[$\S$] A power-law model.\\
 \footnotemark[$\|$]A power-law times exponential cutoff model.\\
\footnotemark[$\#$] Band function.\\
\end{tabnote}
\end{table*}

\end{document}